\documentclass[runningheads]{llncs}

\usepackage{amsmath,amssymb,amsfonts}
\usepackage[dvipsnames]{xcolor}
\usepackage{ifthen}

\newboolean{showcomments}
\setboolean{showcomments}{true}

\ifthenelse{\boolean{showcomments}}
  {
	
	  \newcommand{\nb}[2]{
    		\fbox{\bfseries\sffamily\scriptsize#1}
		{\sf\small$\blacktriangleright$\textit{#2}$\blacktriangleleft$}
   	  }
  }
  {
	\newcommand{\nb}[2]{}
	
}

\newcommand\DDP[1]{\textcolor{Plum}{\nb{Daniele}{#1}}}

 \usepackage[T1]{fontenc}
\usepackage[hidelinks]{hyperref}
\usepackage{cleveref}
\usepackage[numbers]{natbib}
\usepackage{xspace}
\usepackage{url}
\usepackage{comment}
\usepackage{graphicx} 
\usepackage{caption}
\usepackage{subcaption}
\usepackage{orcidlink}

\newcommand{\ie}{\emph{i.e.,}\xspace}
\newcommand{\eg}{\emph{e.g.,}\xspace}

\newcommand{\mas}{$\mathit{MAS}$\xspace}
\newcommand{\ccm}{CoCoME\xspace}

\newcommand{\stp}{ST+\xspace}

\begin{document}

\title{Towards Assessing Spread in Sets of Software Architecture Designs}
\titlerunning{Towards Assessing Spread in Sets of Software Architecture Designs}

\author{
Vittorio Cortellessa\inst{1}\orcidlink{0000-0002-4507-464X}
\and
J. Andres Diaz-Pace\inst{2}\orcidlink{0000-0002-1765-7872}
\and
Daniele Di Pompeo\inst{1}\orcidlink{0000-0003-2041-7375}
\and
Michele Tucci\inst{1}\orcidlink{0000-0002-0329-1101}}
\authorrunning{Cortellessa et al.}
\institute{
University of L'Aquila, Italy
\and
ISISTAN, CONICET-UNICEN, Buenos Aires, Argentina\\
\email{\{vittorio.cortellessa,daniele.dipompeo,michele.tucci\}@univaq.it, andres.diazpace@isistan.unicen.edu.ar}
}

\maketitle              

\begin{abstract}

Several approaches have recently used automated techniques to generate architecture design alternatives by means of optimization techniques.
These approaches aim at improving an initial architecture with respect to quality aspects, such as performance, reliability, or maintainability.
In this context, each optimization experiment usually produces a different set of architecture alternatives that is characterized by specific settings.
As a consequence, the designer is left with the task of comparing such sets to identify the settings that lead to better solution sets for the problem.
To assess the quality of solution sets, multi-objective optimization commonly relies on quality indicators.
Among these, the quality indicator for the maximum spread estimates the diversity of the generated alternatives, providing a measure of how much of the solution space has been explored.
However, the maximum spread indicator is computed only on the objective space and does not consider architectural information (e.g., components structure, design decisions) from the architectural space.
In this paper, we propose a quality indicator for the spread that assesses the diversity of alternatives by taking into account architectural features.
To compute the spread, we rely on a notion of distance between alternatives according to the way they were generated during the optimization. We demonstrate how our architectural quality indicator can be applied to a dataset from the literature.

\end{abstract} 

\section{Introduction}

When designing or evolving software architectures, the improvement of quality attributes like performance, reliability, or maintainability is a central concern for the designer.
This task has recently been the target of an increasing number of automated approaches whose goal is to generate improved versions of an initial architecture~\citep{Cortellessa_Di_Pompeo_Stoico_Tucci_2023,Ni_Du_Ye_Minku_Yao_Harman_Xiao_2021}.
The generated architectures are referred to as design alternatives. These alternatives are automatically generated 
through refactoring, which is the application of transformations for improving some quality attributes of interest while keeping the software functionalities unchanged.
In practical cases, the designer tackles multiple attributes at the same time. For example, she could aim at improving performance while keeping reliability high and the cost of refactoring low. 
This context is ideal for multi-objective optimization.

Multi-objective optimization normally needs a series of experiments to determine appropriate configuration parameters. 
Examples of these parameters can be the choice of the optimization algorithm and its settings, such as the specific genetic algorithm and its population size.
A recurring problem is that of deciding which configuration parameters are to be preferred on the basis of the obtained solution sets.
In multi-objective optimization, this issue has been traditionally addressed by means of quality indicators for solution sets~\citep{LiY19}.
These indicators seek to estimate desirable properties of solution sets and, consequently, can help designers to choose parameters that improve such properties.
Among these properties, the spread of a solution set plays a fundamental role in assessing how much of the solution space was covered and how diverse the solutions are in the set.
The \emph{maximum spread} \emph{(MS)} ~\citep{ZitzlerDT00} is a prominent indicator for such a property.
Higher \emph{MS} values indicate that the optimization searched the solution space enough to cover a wide area.

The spread is especially relevant when dealing with architectural alternatives. It highlights that a sizeable number of alternatives were found, and the obtained solution set covers tradeoffs that are dispersed enough to provide a variety of design choices to the designer.
Note that quality indicators are computed on the problem objectives, which are the attributes being optimized (\ie performance, reliability, and cost in the previous example).
Although these indicators are valuable for assessing the performance of optimization algorithms, they do not provide insights on the architectural features of the solutions, \ie the architectural space, which represent more closely the object of the software engineering task we described.

In this paper, we propose an architectural quality indicator, called \mas \textit{(Maximum Architecture Spread)}, for the estimation of the spread in sets of design alternatives.
\mas is based on the notion of maximum spread by \citet{ZitzlerDT00}, but it operates in the architectural space rather than on the objective space.
To demonstrate our approach, we show how \mas can be applied to a dataset that addresses an architecture optimization problem from the literature. To do this, we calculate the differences among architectural alternatives on the basis of their structure.
This is achieved by first encoding the architectures in terms of sequences of refactorings and then using distance metrics on such encodings.
Our preliminary results indicate that \emph{MAS} provides an additional architectural view of the optimization, and exhibits a correlation with the \emph{(MS)} indicator. \section{Motivating Example}

\begin{figure}
    \centering 
    \includegraphics[width=.98\textwidth]{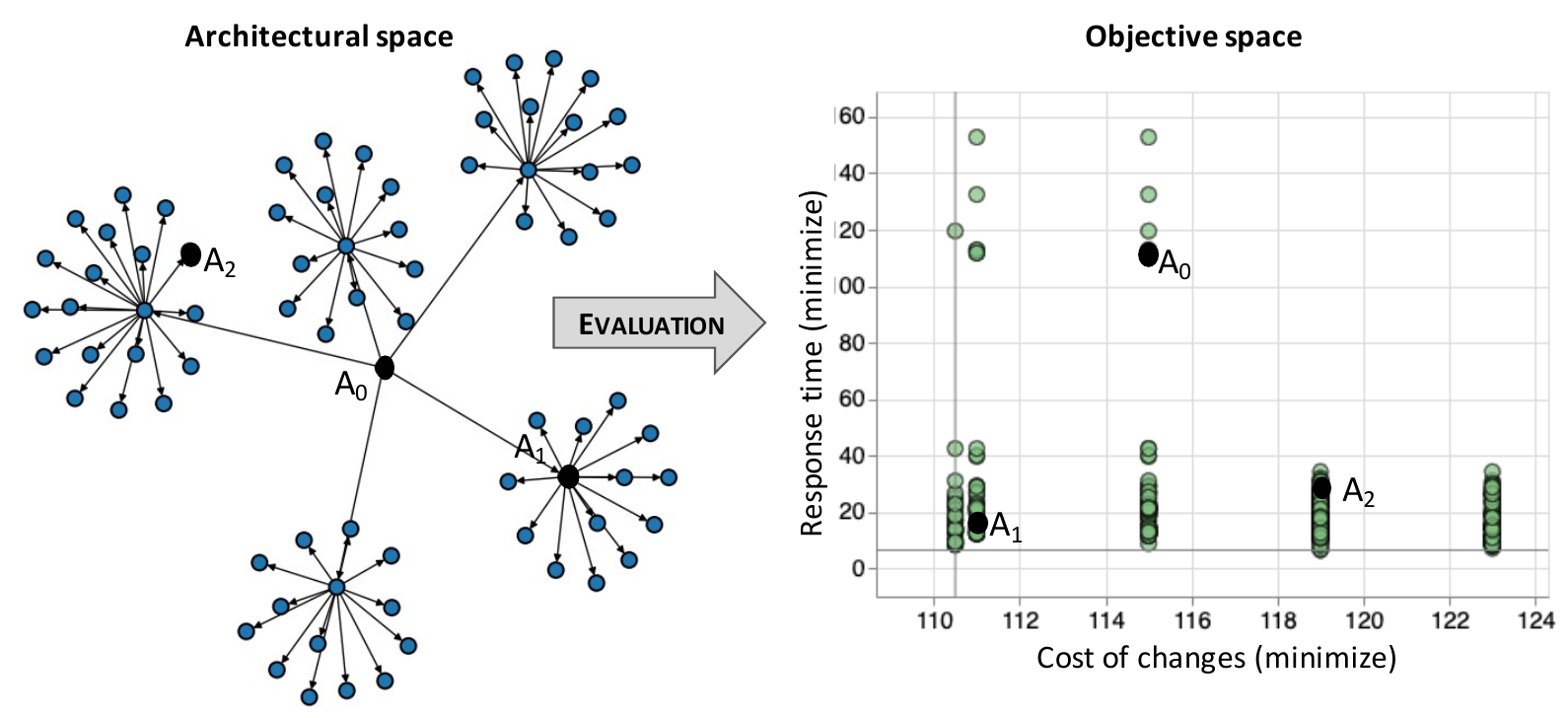}
    \caption{Relationship between the architectural and objective spaces. The architectural space is represented as search tree rooted at the initial architecture ($A_0$). }\label{fig:spaces}
\end{figure}

Let us assume that the designer wants to explore design alternatives that improve a performance objective (\eg minimization of response time) for an initial architecture ($A_0$), while keeping the cost of changes low. For example, this is the case of the \textit{\stp} system~\citep{RagoSBCARS17}, in which architectures are specified in the Palladio Component Model (PCM) notation. 
An automated optimization tool progressively applies refactorings via a heuristic search and generates a set of design alternatives in the \textit{architectural space} (\ie the space of PCM architectures).
This set can be seen as a tree rooted at $A_0$, in which the nodes correspond to alternatives and the edges correspond to refactorings from one alternative to another, as sketched in \Cref{fig:spaces} (left). 
Each alternative is evaluated
to provide quantifiable measures for the objectives. 
Measures for such objectives constitute the \textit{objective space}, in which the designer can analyze tradeoffs between the alternatives, as depicted in \Cref{fig:spaces} (right). 
However, the objective space does not reveal details about the structural characteristics or features of the design alternatives. Likewise, the architectural space does not provide insights about tradeoffs between the objectives. 

To assess the quality of solution sets in the objective space,
the designer can rely on quality indicators, such as \emph{MS}~\citep{ZitzlerDT00}. 
However, analyzing only the objective space can be shortsighted, because it does not expose, for example, architectural characteristics of the PCM alternatives in the solution sets.

We argue that insights from the architectural space are crucial for the designer to understand and compare configuration parameters in order to make the optimization process more useful and efficient. 
Unfortunately, quality indicators for the architectural space have not been reported in the literature, partially because they are not domain independent (as the traditional indicators for the objective space) but depend on the kind of models populating the architectural space. 
We refer to the quality indicators for this space as \textit{architectural quality indicators}. \section{Related Work}\label{sec:related}

We highlight that none of the identified prior studies presented metrics to estimate sets of design alternatives in the architectural space, rather than in the objective space.
\citet{Esfahani_Malek_Razavi_2013} introduced a quantitative framework to graphically drive architects in selecting design alternatives by means of fuzzy mathematical methods to rank alternatives. 
Ranking alternatives helps designers to find the optimal alternative (\ie the best case) and the most critical one (\ie the worst case).
\citet{Sedaghatbaf_Azgomi_2019} proposed a framework for modeling and evaluating software architectures. 
They used a multi-criteria decision model to extract the best and worst alternatives.
To this extent, they introduced a distance metric that extracts the maximum and minimum values for the best and worst alternatives, respectively.
In addition, they support the designer in this complex process by providing a tool named \emph{SQME}.
\citet{Rahmoun_Mehiaoui-Hamitou_Borde_Pautet_Soubiran_2019} exploited a genetic algorithm to generate model transformations and obtain design alternatives defined through the Architecture Analysis and Design Language (AADL). 
To compare the alternatives generated by their approach and the optimal one, the authors introduced a distance metric based on Mixed-Integer Linear Programming (MILP). 
The idea beyond the study by \citeauthor{Rahmoun_Mehiaoui-Hamitou_Borde_Pautet_Soubiran_2019} is to find a way to measure how far the generated alternatives are from the optimal ones.

 \section{Proposed Architectural Quality Indicator}\label{sec:arc-spread}

In multi-objective optimization, the spread of a solution set is recognized as the region of the objective space that is covered by the set.
When the solution set is a Pareto front, the spread is also known as the coverage of the set~\cite{LiY19}.
Higher values of the spread are an indication that the optimization process has extended the search enough to reach a wide area of the objective space.

The \emph{maximum spread (MS)} is a well-known quality indicator \cite{ZitzlerDT00} that measures the range of a solution set by considering, for any two solutions in the set, the maximum extent of each objective.
It is defined as:
\[ \mathit{MS}(S) = \sqrt{ \sum_{i=1}^{o} \max_{s,s^{\prime} \in S}(s_i - s^{\prime}_i)^2 } \]
where $S$ is the solution set under consideration, $s$ and $s^{\prime}$ are solutions in that set, and $o$ is the number of objectives.
Higher values of $MS$ are to be preferred, as they represent a better coverage of the solution space.

As it is evident, quality indicators like $MS$ are defined on the objective space, and they are considered a valuable mean to assess the performance of a search algorithm in producing solution sets coverage.
However, when these optimization techniques are employed in the software architecture context, and specifically for the generation of design alternatives, the quality of the resulting solution sets should be assessed also in the architectural space.
Indeed, the architectural features of the solutions represent the final product of the optimization process, and the designer will make decisions on their basis.
Therefore, we provide an estimate of the coverage and diversity of solution sets in the architectural space.

We introduce the \emph{maximum architectural spread (MAS)} by following the same principles behind the original $MS$.
It is defined as:
\[
\mathit{MAS}(S) =
\sqrt{
  \dfrac{
    \sum_{n=1}^{N} \max_{s,s^{\prime} \in S}( d(s_n, s^{\prime}_n) )^2
  }
  { N \max(d)^2 }
}
\]
where $S$ is the considered set, $N$ is its cardinality, while $s$ and $s^{\prime}$ are solutions in that set.
Moreover, $d(s, s^{\prime})$ is a distance metric that is relevant for the problem, and that can be plugged in the formula to quantify the distance between two architectures.
$max(d)$ is the maximum possible value for the distance metric.
\mas can be interpreted as the square root of the ratio of two quantities: (i) the sum of the maximum distance of any two solutions in the set, divided by (ii) the maximum achievable spread, that is the maximum value for the distance metric multiplied by the number of solutions in the set.
The \mas denominator is used to normalize its value between $0$ and $1$. 
When \mas is $0$, it represents the limit case in which the set consists of a single solution.
Instead, $1$ represents the maximum achievable spread, which occurs when every solution is at maximum distance from another.
The intent is to provide, numerically, an intuition of how far we are from the maximum diversity of solutions we could theoretically achieve.
In addition, the normalization is aimed at enabling the comparison of \mas obtained with different distance metrics.

\subsection{Architectural Distance as Sequence Distance}
The \mas computation depends on having a notion of distance $d(s, s^{\prime})$ between architecture solutions. Since architectures are usually complex objects, simplifying their representations can help to define intuitive distance metrics. In this work, we rely on the sequences of transformations applied to the initial architecture, and use those sequences as proxies for architectural representations. A sequence is actually a path in the search tree, as illustrated in \Cref{fig:spaces} (left). More formally, an architecture $A_i$ is modeled by a sequence of transformations $T_i=<t_{1i},t_{2i},...,t_{Li}>$ of length $L$, which comes from the shortest path between the initial node $A_0$ and the node for $A_i$ in the tree. For each $t_{ki}$, the architectural elements targeted by the transformation are parameters in the representation. 

Once architectures are encoded as sequences, the distance $d(A_i,A_j)$ is defined in terms of the delta of changes between the elements of their respective sequences. If two transformations $t_{ik}$ and $t_{jk}$ share the same name and parameters, then $d(t_{ik},t_{jk})=0$; and conversely, if they have different names and completely different parameters, then their distance is equal to $1$. The problem becomes one of matching sequences, and well-known distances such as \textit{Levenshtein} \cite{Levenshtein:1965} can be used to compute the sequence distances. To distinguish between the transformation names and their parameters, when matching sequences, we use a separate distance function for each part of the transformation, and combine the results via a weighted sum. For a pair of sequences of length $L$, the distance computation is defined as:
\[ \mathit{d}(A_i,A_j) = \sum_{k=1}^{L} simpred(t_{ik}, t_{jk})*w_{pred} + simargs(t_{ik}, t_{jk})*w_{args} \]
Functions $simpred()$ and $simargs()$ extract the transformation name and arguments from $t_{ik}$ and $t_{jk}$, respectively, and then calculate the \textit{Levenshtein} formula. We perform a label encoding of the vocabulary of transformation names and arguments used in the architectural space, before evaluating $simpred()$ and $simargs()$. This encoding maps each transformation name or argument to a unique symbol. The maximum possible values of $simpred()$ and $simargs()$ is $L$. The contributions of these functions are weighted by $w_{pred}$ and $w_{args}$, with the constraint $w_{pred}+w_{args}=1$ to keep $d(A_i,A_j)$ bound to the interval $[0..L]$.
 \subsection{Application of \mas}

We computed \mas on the \textit{ST+} dataset\footnote{\url{https://github.com/SQuAT-Team/paper-supplementary}}, which comprises nine component types for instantiating an architecture alternative, and four quality-attribute scenarios as the optimization objectives. 
Two objectives (referred to as \textit{p1} and \textit{p2}) involve minimizing response time and CPU utilization \cite{PCM2LQN}, while the other two (referred to as \textit{m1} and \textit{m2}) involve minimizing the cost of changes via a complexity metric.
The architectural space contains $554$ candidate architectures.
\begin{figure}
    \centering
    \begin{subfigure}{.49\textwidth}
        \includegraphics[width=\textwidth]{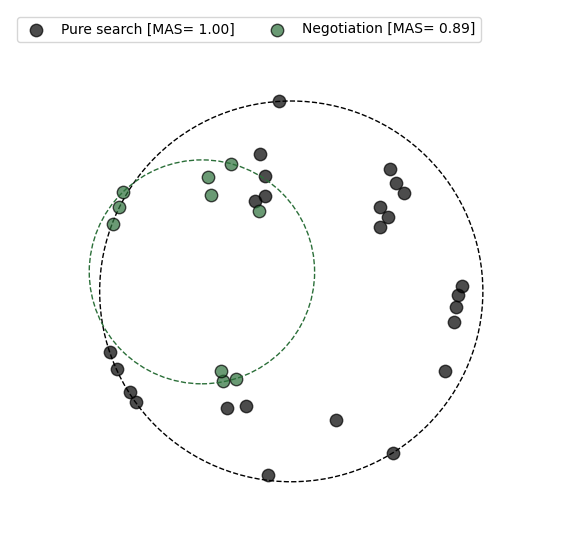}
        \caption{Architectural space}
        \label{fig:stplus_architectural_space}
    \end{subfigure}
    \hfill
    \begin{subfigure}{.49\textwidth}
        \includegraphics[width=\textwidth]{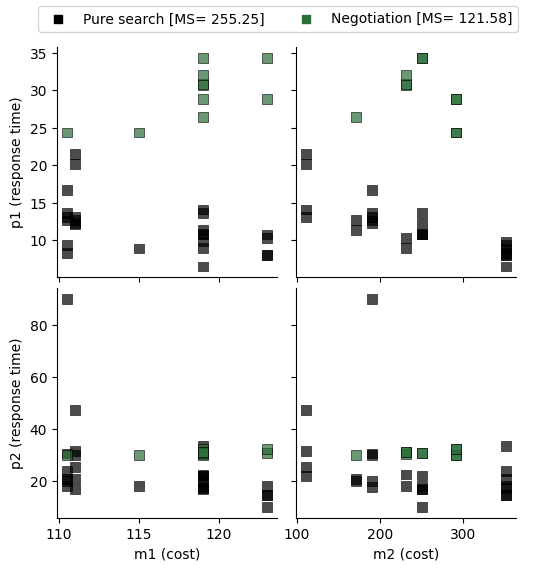}
        \caption{Objective space}
        \label{fig:stplus_objectives_space}
    \end{subfigure}

    \caption{Comparison of the architectural and objective spaces for the two search algorithms in\textit{ ST+}, along with computed \mas and \emph{MS}.}
    \label{fig:stplus}
\end{figure}
Graphical representations of the architectural and quality-attribute spaces are shown in \Cref{fig:stplus}. The colors refer to the two search strategies (standard search and negotiation) exercised for \textit{ST+}. The architectural space (\Cref{fig:stplus_architectural_space}) is visualized using a multi-dimensional scaling (MDS) projection that is derived from the distances $d(A_i,A_j)$ between the PCM architectures. In our MDS chart, each color-coded circle encloses the alternatives returned by an algorithm, and gives a notional view of the spread it achieved.

In the objective space (\Cref{fig:stplus_objectives_space}), the solutions computed by the standard search lie in the Pareto front, while the negotiated solutions set apart, particularly for $p1$ with respect to $m1$ and $m2$. This would mean that the negotiation algorithm tries to balance the utilities for that tradeoff. The \textit{MS} values for the two sets show a larger spread in favor of the standard search algorithm.

In the architectural space (\Cref{fig:stplus_architectural_space}), we observe that the negotiation algorithm achieved a slightly lower \mas than the standard search algorithm, whose spread was maximal. It seems that the space covered by the negotiation algorithm concentrates on a well-defined region. In contrast, the other set is more diverse. Interestingly, some solutions in the negotiation set are at a very close distance of the solutions in the other set. This close distance among the two types of solutions is not evident when looking at the objective space (\Cref{fig:stplus_objectives_space}).

One of the main findings from this experiment is the relationship between the \mas and \textit{MS} indicators. Although they operate on different spaces and assess different objects (\ie architectures and quality values, respectively), there is an apparent correlation between their spreads. 
The observed correlation, however, might have been influenced by the search algorithms used by \textit{SQuAT}. 

If the relationship between \mas and \textit{MS} proves to hold, it might help the designer to make assumptions about the spread on the objective space while looking at the architectural space, and vice versa.  Another observation refers to the architectural distance being used in our evaluation. The assessment of architectural similarities (or dissimilarities) based on their transformation sequences works in the context of an initial architecture and a neighborhood of alternatives that are reachable only via 
transformation sequences applied to the initial architecture.
Therefore, the underlying assumption is that all the sequences have the same starting point. The proposed \textit{Levenshtein} metric, however, is not intended to be a global metric for arbitrary architectures. Furthermore, the metric can be affected by the transformation encoding and by the sequence length. \section{Conclusion}

In this work, we proposed a quality indicator for estimating the spread in sets of software architectures, namely the \emph{maximum architectural spread (MAS)}.
We showed, through a lightweight literature review, that these sets of architectures represent design alternatives and arise in a variety of contexts, especially when multi-objective optimization techniques are employed to generate them.
Differently from existing quality indicators in multi-objective optimization, ours aims at computing an estimate of the solutions spread from an architectural point of view.
This was achieved by encoding the sequences of modifications being applied to generate each architecture, and then using distance metrics to calculate how far apart the architectures are from each other.
Moreover, we showed how our \mas indicator can be applied on practical cases to gain insights on the diversity of architectures and compare optimization settings (\eg algorithms).

The maximum spread provides an idea of the extent of the solution set, but says nothing about the inner shape of the set or the distribution of its solutions.
We intend to overcome this limitation in future work by exploring corner cases in which we obtain similar values of \mas on sets that exhibit contrasting distributions. Furthermore, we will explore complementary indicators to assess properties such as uniformity or convergence within the architectural space.

\newcommand{\SoBigDataITAck}{European Union -- NextGenerationEU -- National Recovery and Resilience Plan (Piano Nazionale di Ripresa e Resilienza, PNRR) -- Project: ``SoBigData.it -- Strengthening the Italian RI for Social Mining and Big Data Analytics'' -- Prot. IR0000013 -- Avviso n. 3264 del 28/12/2021}
\vspace{-1em}
\subsubsection*{Acknowledgments}
Daniele Di Pompeo and Michele Tucci are supported by \SoBigDataITAck. J. Andres Diaz-Pace is supported by the PICT-2021-00757 project, Argentina.

\bibliographystyle{splncs04nat}
\bibliography{bibliography}

\end{document}